\documentclass[reprint, secnumarabic, amssymb, amsmath, nobibnotes, 
     aip, jcp, graphicx]{revtex4-1}
\usepackage{graphicx}		
\begin{document}
\title{Configuration weight function method to solve the many-body 
Schr\"{o}dinger equation.}
\author{V.M.Tapilin}
\email{tapilin@catalysis.ru}
\affiliation {Boreskov Institute of Catalysis, Novosibirsk 630090, Russia}
\date{\today}
\begin{abstract}
A method to solve the Schr\"{o}dinger equation based on the use of constant 
particle-particle interaction potential surfaces is proposed. The many-body 
wave function is presented in configuration interaction form with coefficients -
configuration weight functions - dependent on the total interaction 
potential. A set of linear ordinary differential equations for the 
configuration weight functions was developed and solved for particles in 
a infinite well and He-like ions. The results demonstrate that the method is 
variational and provides upper bound for energy of the ground state; even in 
its lowest two-body interaction potential surfaces approximation, it is more 
accurate than the conventional configuration interaction method and demonstrates 
a better convergence with a basis set increase. For He-like ions one configuration 
approximation with non-interaction electrons functions are used as basis set
the calculated energies are below the Hartree-Fock limit. In three configuration 
approximations the accuracy of energy calculation is close to CI accuracy with 
35 configuration taking into account. Four configurations give the energies 
below CI method and slightly below precise calculation with Hylleraas type wave 
functions.   
\end{abstract}
\maketitle

\section{Introduction}
 M{\o}ller-Plesset perturbation theory and configuration interactions are the 
conventional methods of treating electron-electron correlation in the theory
of atoms and molecules \cite{sherrill:110902}. Unfortunately, both of them due to the presence of
the correlation cusp \cite{kato,King} in the wave function reveal slow convergence 
of electron energy with basis set increasing. At the same time, the fast growth
of computational work which is mainly related to the need for calculation of 
four-index two-electron integrals, places a hard limit the basis set size. 

Density functional theory (DFT) \cite{Hohen, Kohn1965, Kohn1996, Chris} is another approach to 
solve quantum many-body problem. Based on the solution of Kohn-Sham equations \cite{Kohn1965}, it has been successfully 
applied to many problems \cite{Chris}. Unfortunately, the exact form of this functional is 
unknown, and its approximated forms do not always provide the required accuracy, for example, 
in treating systems with strong electron-electron correlations \cite{Ivan, PhysRevB.74.125120, 
Dagotto08072005}. 

All of these give reasons for a search other ways of treating the correlation problem. 
To speed up the convergence, explicitly correlated R12 and F12 methods have been developed over the last 
two decades \cite{Kutzelnigg:1985, Klopper1991583, termath:2002, Noga1992497, noga:7738, Noga2000166,
valeev:3990, klopper:6397, klopper:10890, Valeev2004190, Tenno200456, valeev:244106, shiozaki:044118, 
torheyden:171103, peterson:084102, Yousaf2009303,KlopperInt,doi:10.1021/cr200168z}. Following to Hylleraas 
\cite{Hylleraas29,Hylleraas30} and
Boys and Handy, \cite{Boys309,Boys310} these methods are based on representation 
of the wave function as a production of one-particle wave functions and a correction function
explicitly depending on electron-electron spacing. In R12 the correction function linearly depends on 
electron-electron spacing \cite{Kutzelnigg:1985, Klopper1991583}, in F12 this dependence is exponential
\cite{Tenno200456}. Iterative complement interaction method has been formulated 
in \cite{PhysRevLett.93.030403,PhysRevA.72.062110}. This paper is aimed at developing 
another way of treating the correlation problem presented in.\cite{TapilinJSC,
*TapilinJSC17} The theory is based on the introduction of constant particle-particle interaction potential surfaces. It follows directly from the definition of such surfaces that particle-particle interaction acts along the normal to the surface and, therefore, does not influence particle motion on the surface. Thus this motion can be described by a wave function of independent particles, which results in a new exact representation for many-body wave function and a set of equation to determining it. Further a new form of many-body wave function and equations to find it will be proposed and
applied to particles in a infinite square well and He-like ions. 

\section{Configuration weight functions and equations determining them}
Consider the Schr\"{o}dinger equation of $n$ interacting particles 
\begin{equation} \label{schr}
H\Psi=(H_0+V_{int})\Psi=E\Psi, 
\end{equation}
where $H_0$ is the kinetic energy and external field operator, and $V_{int}$ 
particle-particle interaction operator
\begin{eqnarray}\label{h0}
H_0(\mathbf{R})&=&	\sum_{i=1}^n\left[-\frac{1}{2}\nabla_i^2+V(\mathbf{r}_i)\right]=
-\frac{1}{2}\nabla^2_{\mathbf{R}}+V(\mathbf{R}),\label{hind}\\
	V_{int}(\mathbf{R})&=&\frac{1}{p(\mathbf{R})}=
\sum_{i=1}^{n-1}\sum_{j>i}^{n}\frac{1}{r_{ij}}=
\sum_{i=1}^{n-1}\sum_{j>i}^{n}v_{ij}.\label{vint} 
\end{eqnarray}
Here $\mathbf{R}$ stands for a set of particle coordinates $\mathbf{r}_1,...,\mathbf{r}_n$, 
$r_{ij}=|\mathbf{r}_{i}-\mathbf{r}_{j}|$. 

A constant interaction potential surface $V_{int}(\mathbf{R})=1/p$ selects a subspace of particle coordinates in which particles motion is correlated \textit{ab origin} due to the demand remaining at the surface rather than particle interaction. The resulting interaction force, acting at the interacting particles on the surface, directs along the normal to the surface and does not act on particle movement along the surface, giving rise to redistribution of the particles between surfaces only. Thus, the eigenfunctions of (\ref{h0})
\begin{equation}\label{psi1}
	\Phi_{\mathbf{i}}(\mathbf{R})=\widehat{P}\phi_{i_1}(\mathbf{r}_1)\ldots\phi_{i_n}(\mathbf{r}_n)
\end{equation}
where $\widehat{P}$ is an operator symmetrizing wave function according the system spin, 
 satisfies of (\ref{schr}) on the constant interaction potential surface with eigenvalues
\begin{equation}
\epsilon_{\mathbf{i}}=\epsilon_{i_1}+...+\epsilon_{i_n}+1/p	.
\end{equation}
Here we introduced vectors $\mathbf{i}$ with components $i_1,\ldots,i_n$.
Function (\ref{psi1}) does not satisfy to (\ref{schr}) in the whole
space due to a particle redistribution from surface to surface owing to changing $p$.
We represented the function satisfying (\ref{schr}) in the form
\begin{equation}\label{Psi}
	\Psi(\mathbf{R})=\sum_{\mathbf{i}}\chi_{\mathbf{i}}(p(\mathbf{R}))\Phi_{\mathbf{i}}(\mathbf{R}).
\end{equation}
Function (\ref{Psi}) has the form of configuration interaction function in which coefficients
are replaced by functions $\chi_{\mathbf{i}}(p(\mathbf{R}))$ depending on interaction potential
at points $\mathbf{R}$. This function determines the contributions of 
different configurations for each constant interaction potential surface $p$ and hereinafter 
referred to as \textit{configuration weight function}. Here and below functions $\chi$ and 
$\Phi$ without subscripts mean vector functions with
the components  $\chi_{\mathbf{i}}$ and $\Phi_{\mathbf{i}}$ respectively.

The result of action of Laplace operator at function (\ref{Psi}) can be written as
\begin{equation}
  \begin{split}
	\nabla^2_i&\Psi(\mathbf{R})=\Phi(\mathbf{R})\nabla_i^2\chi(p(\mathbf{R}))+\\
	&2\nabla_i\chi(p(\mathbf{R}))\nabla_i
	\Phi(\mathbf{R})+\chi(p(\mathbf{R}))\nabla_i^2\Phi(\mathbf{R})=\\
	&\Phi(\mathbf{R})(\nabla_i p)^2
	\frac{d^2\chi(p)}{dp^2}+\left[\Phi(\mathbf{R})
	\nabla_i^2p\right.\\
	&\left.+2\nabla_i p\nabla_i\Phi(\mathbf{R})\right]
	\frac{d\chi(p)}{dp}+\chi(p)\nabla_i^2
	\Phi(\mathbf{R}),
	\end{split}
\end{equation}
where results of $\nabla$ operator action on collective 
variable $p$ are 
\begin{align}\label{m2a}
\nabla_ip&=-\frac{1}{V_{int}^2}\nabla_iV_{int}=p^2\sum_{j>i}
\frac{\mathbf{r}_{ij}}{r_{ij}^3},\\ \label{gr_sq}
(\nabla_ip)^2&=p^4\sum_{j>i,k>i}\frac{\cos<\mathbf{r}_{ij},
\mathbf{r}_{ik}>}{r_{ij}^2r_{ik}^2}\nonumber\\&=p^4\sum_{j>i,k>i}
\frac{r_{ij}^2+r_{ik}^2-r_{jk}^2}{r_{ij}^3r_{ik}^3},
\\ \label{grad2}
\nabla^2_ip&=\frac{2}{V_{int}^3}(\nabla_iV_{int})^2-
\frac{1}{V_{int}^2}\nabla^2_iV_{int}=\frac{2}{p}
(\nabla_ip)^2 .
\end{align}
In (\ref{grad2}) it was taken into account that the Coulomb potential is
satisfied to the Laplace equation.

Determining matrix $\mathbf F$ of any operator $\widehat{F}$ on
a surface $p$ by matrix elements 
\begin{equation}
	F_{\mathbf{ij}}(p)=\langle	\Phi_\mathbf{i}|\widehat{F}|
	\Phi_\mathbf{j}\rangle_p=	\int_{\mathbf{S}(p)}d\mathbf{R}
	\Phi_\mathbf{i}\widehat{F}\Phi_\mathbf{j}
\end{equation}
The expression for energy in this notation can be written  in the form
\begin{equation}\label{tener}
	E=\frac{\int dp\chi^{tr}(p)\mathbf{H}(p)\chi(p)}
	{\int dp \chi^{tr}(p)	\mathbf{S}(p)\chi(p)}.
\end{equation}
where $\mathbf{H}$ and $\mathbf{S}$ are the Hamiltonian and overlap matrices,
$\chi^{tr}$ means the transpose of column vector function $\chi$.
It should be noted that functions $\Phi_{\mathbf{i}}$, orthogonal in the whole space, can be
unorthogonal on a surface. Moreover, the set of functions which are linear independent in 
the whole space can became linear dependent on the surface. 

Energy minimization in respect to $\chi_{\mathbf{i}}$ leads to
equations
\begin{equation}\label{eqc}
\begin{split}
	-\frac{\mathbf{T}(p)}{2}\frac{d^2\chi(p)}{dp^2}-\left(\frac{\mathbf{T}(p)}{p}+\frac{\mathbf{U}(p)}{2}\right)\frac{d\chi(p)}{dp}\\ 
  +\left(\mathbf{H_0}(p)+\frac{\mathbf{S}(p)}{p}\right)\chi(p)=E\mathbf{S}(p) \chi(p),
\end{split}
\end{equation}
where
\begin{align}
	 \label{T} T_\mathbf{ij}(p)&=\langle	\Phi_\mathbf{i}
	(\mathbf{R})|(\nabla_i p)^2|\Phi_\mathbf{j}(\mathbf{R})\rangle_p, \\
	 \label{U} U_\mathbf{ij}(p)&=\langle	\Phi_\mathbf{i}
	(\mathbf{R})| 
	\nabla_i p\nabla_i|\Phi_\mathbf{j}(\mathbf{R})\rangle_p ,\\
	 \label{H} H_{0,\mathbf{ij}}(p)&=\langle	\Phi_\mathbf{i}(\mathbf{R})|H_0(\mathbf{R})
	|\Phi_\mathbf{j}(\mathbf{R})\rangle_p \\ 
	 \label{S} S_\mathbf{ij}(p)&=\langle	\Phi_\mathbf{i}(\mathbf{R})|
	\Phi_\mathbf{j}(\mathbf{R})\rangle_p . 
	\end{align}
Matrices $\mathbf{U}(p)$ and $\mathbf{H_0}(p)$ containing $\Phi$ derivatives are nonsymmetric. It is easy to obtain
\begin{equation}
  \mathbf{H_0}(p)=
	\begin{pmatrix}
	\epsilon_1S_{11}(p)&\epsilon_2S_{12}(p)&\cdots&\epsilon_{n_f}S_{1n_f}(p)\\
	\epsilon_1S_{21}(p)&\epsilon_2S_{22}(p)&\cdots&\epsilon_{n_f}S_{n_f}(p)\\
  \hdotsfor[2]{4}\\
	\epsilon_1S_{n_f1}(p)&\epsilon_2S_{n_f2}(p)&\cdots&\epsilon_{n_f}S_{n_fn_f}(p)\\
\end{pmatrix}
\end{equation}
 Obviously after integration over $p$ matrices $\mathbf{S}$ and $\mathbf{H}_0$ become diagonal with matrix elements $S_{ij}=\delta_{ij}$ and $H_{0;ij}=\epsilon_i\delta_{ij}$, 
$U$ becomes symmetric. Matrices $\mathbf{H_0}(p)$ and $\mathbf{H}_0^{tr}(p)$ have the
same eigenvalues $\epsilon_i$. It should be note that artificial symmetrization of $\mathbf{H_0}(p)$
by sum of $\mathbf{H_0}(p)$ and $\mathbf{H}_0^{tr}(p)$ leads to incorrect results even for non-interaction particles.  

Eq. (\ref{eqc}) is a set of linear ordinary differential equation with eigenvalues
equal to the system energy. The terms containing derivatives of $\chi$ describe 
additional contributions to kinetic energy, arising when redistribution 
of electrons between different interaction potential surfaces occurs. There is no 
such redistribution for non-interacting particles. For this case functions (\ref{psi1}) are
eigenfunctions of $\mathbf{H}$, functions $\chi(p)$ do not depend on $p$ and differential equations 
(\ref{eqc}) reduce to the Schr\"{o}dinger equations of non-interacting particles.  
Due to asymmetry of matrices in (\ref{eqc}) besides configuration
weight function $\chi(p)$ there is another set of configuration function $\chi^l$ (superscript $l$
means 'left'), which is a solution of Eq. (\ref{eqc}) with transposed matrices.
 
The boundary conditions for $\chi$ follow from the demand for $\Psi$ to be finite 
in the whole space. At $p=0$ at least two particles are at the same space point.
In the neighborhood of such points, as it was shown in\cite{kato},  the wave function 
behaves as $e^{r_{12}/2}$, which means that the 1st derivative of the wave function
is discontinuous at such points, giving rise to the cusp problem - slow convergence
of the wave function to the exact one with increasing the basis set. Consistent with 
our theory there is no cusp problem at all because points of the wave function discontinuity 
are lying at the boundary point. To find the boundary conditions at $p=0$ rewrite 
(\ref{gr_sq}) in the form
\begin{equation}\label{g0}
	(\nabla_ip)^2=p^4\left(\sum_{j>i}\frac{1}{r_{ij}^4}+\sum_{k>i}
	\frac{\cos<\mathbf{r}_{ij},\mathbf{r}_{ik}>}{r_{ij}^2r_{ik}^2}\right)
\end{equation}
and expand (\ref{g0}) into the Taylor series. The expansion can be performed for two, three
etc. particles at the same point, however, in all cases the results will be the same
\begin{equation}\label{g0a}
	(\nabla_ip)^2=1+O(r_{12})
\end{equation}
so when $p\rightarrow 0$ (\ref{eqc}) reduced to 1st order differential equations
\begin{equation}\label{eqc0}
\begin{split}
	-\frac{\mathbf{S}(p)}{p}\frac{d\chi(p)}{dp}+\frac{q\mathbf{S}(p)}{p}
\chi(p)=0,
\end{split}
\end{equation}
with a restricted at $p=0$ solutions
\begin{equation}\label{sol0}
\begin{split}
\chi_i(p)=\chi_i(0) e^{qp/2}
\end{split}
\end{equation}
The wave function behavior (\ref{sol0}) coincide with $e^{r_{12}/2}$ presented in paper\cite{kato}.  Eqs. (\ref{sol0}) provide us with the bounder conditions at $p=0$.

For $p\rightarrow\infty$ the interaction vanish and according to (\ref{m2a})-(\ref{grad2})
matrices $\mathbf{S}^{-1}\mathbf{T}$, $\mathbf{S}^{-1}\mathbf{U}$, and 
$\mathbf{S}^{-1}\mathbf{H}_0$ tends to constant, so a solution of
(\ref{eqc}) can be approximated at a point $p$ as $e^{\lambda p}$. 
Substitution of this representation in ($\ref{eqc}$) leads to
\begin{equation}\label{eqp}
\begin{split}
	\sum_{j=1}^{n_f}\left[-\lambda^2T_{ij}(p)/2
	-\lambda ((T_{ij}(p)/p+U_{ij}(p))+H_{0;ij}(p)\right.\\ \left.
	+S_{ij}(p)/Zp-ES_{ij}(p)\right]\chi_j(p)=0,\ i=1,...,n_f,
\end{split} 
\end{equation}
$n_f$ is the number of configurations taken into account.
 Set (\ref{eqp}) has non-zero solution if
\begin{equation}\label{deti}
\det(\Lambda)	=0
\end{equation}
where matrix $\Lambda$ is determined by the expressions in the square brackets of (\ref{eqp}). 
Obviously, $\det(\Lambda)$ is a $2n_f$ order polynomial of $\lambda$. The
$2n_f$ roots of the polynomial, possibly complex, will be denoted $\lambda_i$. Not all of
the roots have physical meaning. The demand that wave function must be finite in the whole
space leads to
\begin{equation}\label{fncp}
	\chi_i(p)S_{ij}\chi_j(p)\approx e^{\lambda_i(p)+\lambda_j(p)}S_{ij}<\infty
\end{equation}
Another restriction on the choice of physical meaning $\chi_i(p)$ results from demand that
with the switch off the particle-particle interaction $\chi_i(p)$ becomes constant, so
$\lambda_i(p)\rightarrow 0$ when $q\rightarrow 0$. As a result, the number of  
physical meaning $\lambda_i(p)$ does not exceed $n_f$. The demand (\ref{fncp}) provide us the
with bounder condition for (\ref{eqc}) for a big $p$.

\section{Constant interaction potential surfaces and their approximations}
\begin{figure}
	\centering
		\includegraphics[scale=0.3]{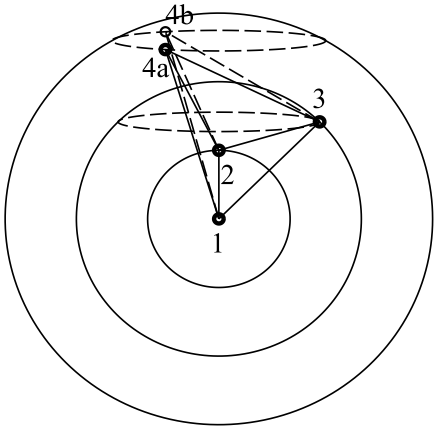}
	\caption{A four particle tetrahedron. Particles (small numerated 
circles) are placed at the framework vertexes, the length of the framework edges 
equals to $r_{ij}$. Solid circles are sphere cross-section by $x=0$ plane, dash curves are circles obtained by crossing spheres by $z_i=r_{1i}\cos\upsilon_{1i}$ planes, 4a and 4b points correspond to two points in (\ref{xyz})}.  
	\label{f:Graph1}
\end{figure}

In case of two particles the constant interaction potential surface is a sphere of radius $r_{12}$ with the center at the position of the selected particle. In case of n-particles the values $r_{1i}$, $i=2,\ldots ,n$, determine $n-1$
spheres remaining on which particles do not change the interaction potential with the first one; even the total potential will change. Setting $r_{2i}$, $i=3,\ldots ,n$, to save interaction potential with the first two particles the rest of the particles must move along the circles of radius $\rho_i=r_{1i}\sin\upsilon_{1i}$ obtained by the crossing lines of the spheres and planes $z_i=r_{1i}\cos\upsilon_{1i}$, $\cos\upsilon_{1i}=(r_{1i}^2+r_{12}^2-r_{2i}^2)/2r_{12}r_{1i}$, in a coordinate system with $z$-axes directed along $\mathbf{r}_{12}$. The values $r_{2i}$ are not arbitrary but must satisfy the conditions
\begin{equation}\label{ineq1}
 |r_{1i}-r_{12}|\le r_{2i}\le r_{1i}+r_{12}. 
\end{equation}
Setting $r_{3i}$, $i=4,\ldots , n$ determines two points at the circles with coordinates 
\begin{equation}\label{xyz}
	z_i=r_{1i}\cos\upsilon_{1i},\ y_i=\rho_i\cos\varphi_i, \ x_i=\pm\rho_i|\sin\varphi_i|.
\end{equation}
where $\cos\varphi_i=(\rho_i^2+\rho_3^2+(z_i-z_3)^2)/2\rho_i\rho_3$.	
 The values of $r_{3i}$ must satisfy to inequalities 
\begin{equation}\label{ineq2}
(\rho_i-\rho_2)^2+(z_i-z_2)^2\le r_{3i}^2\le (\rho_i+\rho_2)^2+(z_i-z_2)^2.
\end{equation}
 The set of $r_{ij}$ with $i\le 3$ and $j>i$ values determines a solid polyhedron with the particles at its vertexes, which is a point at a constant interaction potential surface, shown for four particles in Fig.\ref{f:Graph1}. 
Rotation of the polyhedron around of $\textbf{r}_{12}$ axes, rotation of the axes around $\textbf{r}_1$ point and 
the move of the point in the space determine a constant interaction potential surface. The averaging of any one-body operator $\widehat{F}$ along the surface can be expressed as
\begin{equation}
\begin{split}
	F(p)=\int d\mathbf{r}_1\int r_{12}^2\sin\upsilon_{12}d\varphi_2 d\upsilon_{12}\int r_{13}\sin\upsilon_{13}d\varphi_3\\
	\Phi(\mathbf{r}_1,\ldots ,\mathbf{r}_n)\widehat{F}\Phi(\mathbf{r}_1,\ldots ,\mathbf{r}_n)
\end{split}
\end{equation}
 where $\mathbf{r}_i$ is determined by (\ref{ineq1})-(\ref{ineq2}). So, $3(n-1)$ values $r_{ij}$ with $i\le 3$ and $j>i$, definitely determine the value of particle-particle interaction potential.

There are different combinations of $r_{ij}$ with $i\le 3$ and $j>i$ giving the same value of the interaction potential. As one can see from (\ref{vint}), the constant electron-electron interaction potential surface is a plane in the space of the pair potentials $\{v_{ij}\}$, which will be refered to as \textit{v-space}. The usual space in which particles move will be refer to as \textit{r-space}. The dimensionality of v-space is $n(n-1)/2$, however the restrictions introduced in the above paragraph reduced it to $3(n-1)$. Any point on a constant interaction potential surface $v$ can be moved to surface $v'$ by coordinate scaling
\begin{equation}\label{scl}
r_{ij}(v')=r_{ij}(v)v/v'
\end{equation}
It means that it is enough to construct only one surface, for example,
\begin{equation}\label{pln}
v(v_{12},...,v_{N-1,N}) =1
\end{equation}
and obtain the other ones by scaling transformation (\ref{scl}).

 Each point of v-space determines the relative particle positions $r_{ij}$ in the r-space, so 
the set of $\mathbf{R}$ belonging to the same surface can be easily determined. 
However, due to the multidimensionality of the plane and disability to integrate over 
the variables on the plane independently a numerical integration over the surface can be 
performed only for several particle systems. It means that in practice the developed theory 
can be applied only for such systems, and an extension of the theory to 
bigger systems needs to be simplified. Possible simplifications are proposed below.

At first, a constant potential surface for potential acting on a particle from the 
other ones can be introduced.
For one particle at $\mathbf{r}_1$ it consists of $n-1$ spheres of radius $r_{1i}$. Separate this
potential from the total one 
\begin{equation}\label{pt}
\frac{1}{p_1}=\sum_i\frac{1}{r_{1i}},	
\end{equation}
and determining matrix elements of a operator $\widehat{F}$
\begin{equation}\label{p1n}
 \begin{split}
F(r_{12},...,r_{1n})=\int d\mathbf{r}_1\int d\Omega_{12}...\int d\Omega_{1n}	\\
\Psi^*(\mathbf{r}_1,\mathbf{r}_1+\mathbf{r}_{12},...,\mathbf{r}_1+\mathbf{r}_{1n})\widehat{F}
\Psi(\mathbf{r}_1,\mathbf{r}_1+\mathbf{r}_{12},...,\mathbf{r}_1+\mathbf{r}_{1n}),
 \end{split}
\end{equation}
where $d\Omega_{1i}=r_{1i}^2\sin(\theta_{1i}) d\theta_{1i} d\varphi_{1i}$, and integration over
$\Omega_{1i}$ can be performed independently. As a result, the dimension of the constant potential 
surface becomes $n-1$.

Another possible way of such simplification is an introduction of a set of approximations to the
theory based on the further lowering the dimension of interaction potential surface by averaging
over the moving of a part of particles. Obviously, the averaging over all particles
but one leads to Hartree-Fock approximation. The next approximation - averaging over 
all particles but two - describes the motion of exactly correlating particle pair in
the middle field of other particles can be called \textit{independent pair} approximation. The
same way can be introduced \textit{independent triplet}, \textit{quadruple}, etc. approximations.
As a result, one can obtain the set of equations (\ref{eqc}) in which matrix elements
are calculated as 
\begin{equation}\label{Ft}
F_{\mathbf{ij}}(p_m)=\int_{S(p_m)}d\mathbf{R}_m\int d\mathbf{R}_{n-m}
	\Phi_\mathbf{i}(\mathbf{R})\widehat{F}\Phi_\mathbf{j}(\mathbf{R}),
\end{equation}
where $1/p_m$ denotes potential and $S(p_m)$ is a constant interaction potential surface in 
the space of $m$ particles. Integration over the rest $n-m$ particles can be performed 
independently for each particle coordinates. Operator $\widehat{F}$ can be 
represented as a sum of operators acting in space of $m$ and $n-m$ 
particles, and interaction operator between these two spaces
\begin{equation}
	\widehat{F}(\mathbf{R})=\widehat{F}(\mathbf{R}_m)+
	\widehat{F}(\mathbf{R}_{n-m})+\widehat{F}(\mathbf{R}_m,\mathbf{R}_{n-m}).
\end{equation}
In accordance with this division, the total energy (\ref{tener}) can be
represented as a sum of energies of $m$ and $n-m$ particle systems and 
interaction energy between them. Obviously, energy of $n-m$ particle
system does not take into account particle correlation. It gives the
constant contribution in eigenvalues of (\ref{eqc}). The interaction  
here plays the role of an external field acting on $m$-particle system. 
Thus (\ref{eqc}) is reduced to a set of equations
for $m$-particle in the external field and the middle field of other 
particles. The solutions of (\ref{eqc}) gives an exactly correlated function 
for $m$ particles in the environment described above.
  
\section{Simple exactly solvable examples.}
To test the theory we considered two simple models with directly
solvable Scr\"{o}dinger equation and solved the equation directly, with configuration
interaction method, and with different approximations of the developed theory.
We considered two and three particles in a one dimensional infinite square potential well. 
To avoid errors in derivatives approximation by finite differences and to reduce the numerical 
calculations, we \textit{ab origin} will use the discrete space. The model makes it possible to
solve the Schr\"{o}dinger equation directly. The comparison results obtained on the basis of the
developed methods with the exact ones allows us to estimate the validity and efficiency of the theory. 

\begin{table*}[t]
\vspace{5mm}
 Table I. Energies of the ground and selected exited states of two particles in the 
infinite well for 1, 2, 3, and 8 $\psi$ functions taking into account.	SRF and CI
columns contains the energies obtained with (\ref{chieq}) and (\ref{cieq}) equations,
correspondingly. Relative errors are presented in the brackets.
	\begin{center}
		\begin{tabular}{cccccccc}\hline\hline
state&\multicolumn{2}{c}2 &\multicolumn{2}{c}3&\multicolumn{1}{c}4& 8\\
     &CWF&CI&CWF&CI&CI&CWF,CI,ext.\\ \hline
2&  0.905666(4.$10^{-4}$)&0.927756(2.10$^{-2}$)&0.905284(2.10$^{-6}$)&0.910008(5.10$^{-3}$)&
0.905968(8.10$^{-4}$)&0.905282\\
3&          &        &1.511632(5.$10^{-4}$)&1.536781(2.$10^{-2}$)&1.519998(6.$10^{-3}$)&1.510904\\
4&  1.877636(3.$10^{-2}$)&        &1.821044(6.$10^{-5}$)&1.853396(2.$10^{-2}$)&1.825807(3.$10^{-3}$)&1.820939\\
5&          &        &2.217694(3.$10^{-4}$)&        &2.248223(1.$10^{-2}$)&2.216982\\
6&          &        &2.636500(4.$10^{-2}$)&        &2.564415(2.$10^{-2}$)&2.523710\\
8&  3.065297(1.$10^{-2}$)&        &3.033641(3.$10^{-3}$)&        &3.060787(1.$10^{-2}$)&3.023590\\
13& 4.355834(2.$10^{-3}$)&        &4.358194(1.$10^{-3}$)&        &        &4.362801\\
21& 5.644822(3.$10^{-3}$)&        &5.696051(6.$10^{-5}$)&        &        &5.664018\\
25& 6.824136(8.$10^{-3}$)&        &6.709824(7.$10^{-5}$)&        &        &6.710304\\
28& 7.779231(7.$10^{-5}$)&        &7.779764(1.$10^{-6}$)&        &        &7.779772\\
\hline \hline
			\end{tabular}
			\end{center}
\end{table*}
\subsection*{The model Schr\"{o}dinger equation}
Represent the kinetic energy operator $h$ acting at particle $\alpha$ as
	 \begin{align}\label{hkin}
	 		h(x_\alpha)&=-\frac{d^2\psi_i(x_\alpha)}{d x_\alpha^2}\nonumber\\
		&=\frac{2\psi_i(x_\alpha)
		-\psi_i(x_\alpha+\delta x_\alpha)-\psi_i(x_\alpha-\delta x_\alpha)}{\delta x_\alpha^2}
   \end{align}
	where $x_\alpha$ numerates the points in the well. Lets $m$ is the number
	of such points. 
	The eigen functions $\psi_i(x_\alpha)$ of operator (\ref{hkin}) 
	vanish at the boundary points of a infinite well  are
\begin{align}\label{onef}
\psi_i(x_\alpha)=&\sqrt{\frac{2}{\pi}}\sin{ix_\alpha},\:
x_\alpha=j\delta x_\alpha,\:\delta x_\alpha=\pi/(m-1),\nonumber\\
i=&1,\ldots,m-2,\:j=0,\ldots,m-1,
\end{align}
Functions (\ref{onef}) are a complete set of functions in the well.
The Schr\"{o}dinger equation for $n$ particles in the well can be approximated as
\begin{equation}\label{ddh}
	H\Psi=\sum_\alpha^n[h(x_\alpha)+\sum_{\beta=\alpha+1}^nv(x_\alpha, x_\beta)]\Psi=
E\Psi
\end{equation}
where interaction potential between particles $\alpha$ and $\beta$ 
was choose in the form 
\begin{equation}\label{fp}
v(x_\alpha x_\beta)=\frac{q}{|x_\alpha-x_\beta|+\lambda}
\end{equation}
where $\lambda$ is added to interaction to avoid infinity when
$x_\alpha=x_\beta$, $q$ is particle's charge. 
The order of this set of equations is $m^n$, so for two and three particles in the well
the orders are 64 and 512 correspondingly and solution of (\ref{ddh}) can be obtained
by direct diagonalization of matrix $H$. This solution will be a reference point in 
estimating the accuracy of approximated method to solve (\ref{ddh}).

A solution of (\ref{ddh}) $\Psi$ can be represented by a linear combination
of configuration functions $\Phi_\mathbf{i}$
\begin{equation}\label{ci}
	\Psi(\mathbf{x})=\sum_\mathbf{i}c_\mathbf{i}\Phi_\mathbf{i}(\mathbf{x})
\end{equation}
where $\mathbf{i}$ and $\mathbf{x}$ are n-dimensional vectors with components $i_1,\ldots,i_n$ and
$x_1,\ldots,x_n$, correspondingly, and
$\Phi$ is a production of one-body functions symmetrized with operator $\widehat{S}$  
\begin{equation}\label{wf}
	\Phi_\mathbf{i}(\mathbf{x})=\widehat{S}\psi_{i_1}(x_1)\psi_{i_2}(x_2)\ldots\psi_{i_n}(x_n)
\end{equation}
where $\psi_i$ is one particle functions. If $\psi_i$ is a complete set of functions, (\ref{ci}) is 
an exact representation of the wave functions, in other cases (\ref{ci}) only approximates $\Psi$.
Below we have compared the convergence of the approximated functions of configuration interaction (CI)
and configuration weight function (CWF) methods. 

In CI method (\ref{ddh}) is transformed to
\begin{equation}\label{cieq}
	\sum_{\mathbf{i}_k}H_{\mathbf{i_k}\mathbf{j_k}}c_\mathbf{j_k}=Ec_\mathbf{i_k}
\end{equation}
where $\mathbf{i}_k$ and $\mathbf{j}_k$ are $k$-dimensioned vectors containing indexes only 
$k\leq m$ functions $\psi_i$, and
\begin{equation}
	H_{\mathbf{i}\mathbf{j}}=\left\langle\Phi_\mathbf{i}|H|
	\Phi_\mathbf{i}\right\rangle
\end{equation}
 In CWF method the wave function has the form (\ref{ci}),
but coefficients $c_\mathbf{i}$ are replaced by configuration weights functions 
$\chi_\mathbf{i}(p)$ which are dependent on the value of interaction potential 
\begin{equation}\label{srf}
	1/p=\sum_{\alpha,\beta>\alpha}^nv(x_\alpha, x_\beta)
\end{equation}
and the weight functions satisfy to a set of equation
\begin{equation}\label{chieq}
	\sum_{\mathbf{i}_k}[H_{\mathbf{i}_k\mathbf{j}_k}(p)-ES_{\mathbf{i}_k\mathbf{j}_k}(p)]\chi_{\mathbf{j}_k}(p)=0
\end{equation}
  where
	\begin{align}\label{hchi}
	H_{\mathbf{i}_k\mathbf{j}_k}(p)=&\left\langle\Phi_{\mathbf{i}_k}|H|\Phi_{\mathbf{j}_k}\right\rangle_p\\ \label{ochi}
	S_{\mathbf{i}_k\mathbf{j}_k}(p)=&\left\langle\Phi_{\mathbf{i}_k}|
	\Phi_{\mathbf{j}_k}\right\rangle_p
	\end{align}
	and summation is performed only over points $x_\alpha$ and $x_\beta$ satisfying the condition
	(\ref{srf}) for a given $p$. Overlap matrix $S_{\mathbf{i}_k\mathbf{j}_k}(p)$ is appearing because functions 	$\Phi_{\mathbf{i}_k}$, orthogonal in the whole space, become non-orthogonal on subspaces determined 
	by value $p$.	As a results, the set of functions $\Phi_{\mathbf{i}_k}$ linear independent in the whole space 	
	may become linear dependent on a surface. In this case some of eigenvalues of matrix 
	$S_{\mathbf{i}_k\mathbf{j}_k}(p)$ are equal to zero and we reduced the basis function set
	for these surfaces to exclude zero eigenvalues of the matrix.
	
	Equations (\ref{chieq}) are a representation in the discrete space of the equations (\ref{eqc})
Indeed, kinetic energy operator acting in a discrete space at a product of the
functions
\begin{equation}
 \begin{split}
	\frac{\partial \chi(p(\mathbf{x}))\Phi(\mathbf{x})}{\partial x_\alpha}=
	[\chi(p(\mathbf{x}_\alpha,x_\alpha+\delta x_\alpha)
	\Phi(\mathbf{x}_\alpha,x_\alpha+\delta x_\alpha))\\
	-\chi(p(\mathbf{x}_\alpha,x_\alpha-\delta x_\alpha)
	\Phi(\mathbf{x}_\alpha,x_\alpha-\delta x_\alpha]/(2\delta x_\alpha)
 \end{split}
\end{equation}
tends with $\delta x_\alpha\rightarrow 0$ to
	\[\frac{\partial \chi(p(\mathbf{x}))}{\partial x_\alpha}\Phi(\mathbf{x})
	+\chi(p(\mathbf{x}))\frac{\partial \Phi(\mathbf{x})}{\partial x_\alpha}
\]
 
It should be note that kinetic energy operator is a hermitian operator for functions (\ref{wf}) in the whole space
because the wave functions vanish at bounder points\cite{Landau} 
and remains hermitian on the constant interaction potential surfaces for the same reason.  

	Equations (\ref{ddh}), (\ref{cieq}) and (\ref{chieq}) have been solved for two and three particles in the well 	and the results are presented below. 
\subsubsection*{Two particles in a infinite potential well}
\begin{figure}
	\centering
		\includegraphics[scale=0.3]{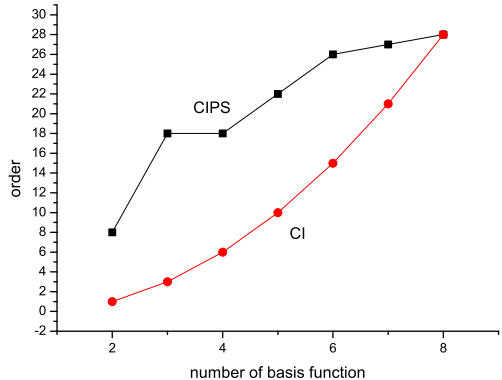}
	\caption{Matrix order for CWF and CI}%
	\label{fig:sysn}
\end{figure}

Energies of ground and some of excited antisymmetric stations for two particles in the wall are
presented in Table I. States are numerated in compliance with the state numeration of $H$ matrix. 
The results obtained with (\ref{cieq}) and (\ref{chieq}) for $k=8$ coincide with the exact ones 
obtained by direct diagonalization of matrix $H$. The number of states obtained with CI method is 
equal to the antisymmetric functions $n_c=k(k-1)$ which can be constructed with $k$ one-body function. 
For CI states presented in Table I the corresponding states of CWF are also shown. However, the
number of exited states calculated with CWF is grater than $n_c$ because $n_c$ states can be 
constructed for each constant interaction potential surface. Not all such constructed functions 
are linear independent which is revealed by appearing of zero eigenvalues of overlap 
matrix (\ref{ochi}). In the calculations the number $n_c$ has been reduced until all the eigenvalues 
become grater than zero. The growth of the number of linearly independent function with the increase 
the number of basis functions for CWF and CI are shown in Fig.\ref{fig:sysn}. The additional to CI 
exited states in Table I were chosen arbitrarily and show the accuracy of the calculated excited 
states  energies.

Table I shows that the CWF relative energy error for $k=2$ two order is less than the CI error. To reach
comparable accuracy CI method needs $k=4$ whereas CWF gives practically exact energy for $k=3$. 
Table I does not present the CWF results for $k=4$ because they are coincide with those for $k=3$ due to 
the equal number of independent configuration for these cases as it can be seen in Fig.\ref{fig:sysn}. 
The accuracy in energies of exited states at first drops with energy; than it starts to grow and 
at the end reaches the accuracy for the ground states.    

Different convergence of CWF and CI methods reflects the different growth in the number 
of operated functions of these method. As one can see in Fig.\ref{fig:sysn} for CWF the number 
of functions grows fast at the beginning and slows at the end wherease CI method shows a slow 
increase at the beginning and fast increase at the end.
\subsubsection*{Three particles in a infinite potential well}
The main aim of solving a three particle model problem is to check the efficiency of different
approximation to exact interaction potential surfaces proposed in Part I. Just as in 
the previous section we solved the problem directly, by CI and CWF methods.  

\begin{figure}
\centering
\begin{picture}(100,70)(50,0)\label{fig.pnt}
 \multiput(0,0)(50,0){5}{\circle*{5}}
 \multiput(0,50)(50,0){5}{\circle*{5}}
 \multiput(100,0)(50,0){3}{\circle{10}}
 \multiput(100,50)(50,0){2}{\circle{10}}
 \put(0,50){\circle{10}}
  \put(0,60){3}
	\put(200,10){3}
  \put(100,60){1}
	\put(100,10){1}
  \put(150,60){2}
	\put(150,10){2}
\end{picture}
\caption{Distinguish between surfaces.
Numerated circles represent partcles}%
\label{fig:pnt}
\end{figure}
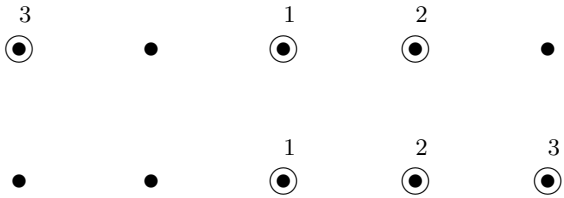

According to the value of the total interaction potential (\ref{fp}), all 
particle coordinate combinations for three particles in the well can be divided into twenty groups
which we called constant interaction potential surfaces and designated below as $p_t$.
Besides, we determined two other interaction potential surfaces, $p_1$ containing 36 surfaces and  
$p_2$ containing 8 surfaces. These surfaces approximate the interaction potential as
\begin{align}
	1/p_1&=1/|x_1-x_2|-1/|x_1-x_3|\\
	1/p_2&=1/|x_1-x_2|
\end{align}
Obviously, on $p_1$ surfaces the interaction potential acting at the first particle is
a constant, on $p_2$ surfaces the interaction potential does not depend on the position
of the 3rd particle. Differences between these types of surfaces is illustrated in 
Fig.\ref{fig:pnt}. Two points' locations presented in Fig.\ref{fig:pnt} determine two 
different surfaces for $p_t$ because they have different value of $|x_2-x_3|$; 
the same surface for $p_1$ because $|x_1-x_2|$ and $|x_1-x_3|$ is the same for both locations. 
These locations belong to the same surface for $p_2$ also; moreover, the change of the 3rd 
particle location does not lead to the change of the surface.  

The exact diagonalization of matrix $H$ gives 120 states belonging to the pure symmetric representation 
of the permutation group (not suitable for electrons), 56 to the pure antisymmetric,  and 336 to 
the mixed symmetric (neither pure symmetric nor pure antisymmetric) representation of the permutation group. 

The energies of the four lowest tates obtained with CI method, and with different approximations of CWF for different $n_f$ are shown in Table II.

\begin{table}
\vspace{5mm}
Table II. Convergence with basis set increase for ground (g) and
	1st, 2nd, and 3rd excited states obtained with $p_t$, 
	$p_1$, $p_2$ and $CI$ matrices.	
	\begin{center}
		\begin{tabular}{cccccc}\hline
$n_f$&state & $p_t$    & $p_1$    & $p_2$ &  CI\\ 
2&g    &  2.5167521 &  2.5523628 & 3.2802597&  4.7233508 \\
 &1st  &  2.6007468 &  2.5630105 & 3.3433421&  4.7233508 \\
 &2nd  &  2.6007468 &  2.5810447 & 3.5953573&  4.9761493 \\
 &3rd  &  2.8993370 &  2.7509650 & 3.7269786&  4.9761493 \\
3&g    &  2.5164929 &  2.5164929 & 2.6297927&  2.7116594 \\
 &1st  &  2.5624931 &  2.5512484 & 2.6779104&  3.1982957 \\
 &2nd  &  2.5624931 &  2.5513833 & 2.6926681&  3.1982957 \\
 &3rd  &  2.6311803 &  2.6246885 & 2.7596655&  3.5842098 \\
4&g    &  2.5164929 &  2.5164929 & 2.5796667&  2.6802123 \\
 &1st  &  2.5512299 &  2.5512296 & 2.5965684&  2.6803017 \\
 &2nd  &  2.5512299 &  2.5512296 & 2.6583777&  2.6803017 \\
 &3rd  &  2.6246904 &  2.6246885 & 2.6678403&  2.7694084 \\
5&g    &  2.5164929 &  2.5164929 & 2.5560362&  2.6101078 \\
 &1st  &  2.5512296 &  2.5512296 & 2.5705697&  2.6273272 \\
 &2nd  &  2.5512296 &  2.5512296 & 2.6028451&  2.6273272 \\
 &3rd  &  2.6246885 &  2.6246885 & 2.6369918&  2.6596377 \\
6&g    &  2.5164929 &  2.5164929 & 2.5434281&  2.5722193  \\
 &1st  &  2.5512296 &  2.5512296 & 2.5623772&  2.5938677 \\
 &2nd  &  2.5512296 &  2.5512296 & 2.5847363&  2.5938677 \\
 &3rd  &  2.6246885 &  2.6246885 & 2.6308173&  2.6415573 \\
7&g    &  2.5164929 &  2.5164929 & 2.5304009&  2.5478066 \\
 &1st  &  2.5512296 &  2.5512296 & 2.5569004&  2.5751546 \\
 &2nd  &  2.5512296 &  2.5512296 & 2.5676064&  2.5751546\\
 &3rd  &  2.6246885 &  2.6246885 & 2.6269072&  26336853 \\
8&g    &  2.5164929 &  2.5164929 &  2.516492&  2.51649929\\ 
 &1st  &  2.5512296 &  2.5512296 &  2.551229&  2.55122966 \\
 &2nd  &  2.5512296 &  2.5512296 &  2.551229&  2.55122966 \\
 &3rd  &  2.6246885 &  2.6246885 &  2.624688&  2.62468855 \\
			\end{tabular}
			\end{center}
\end{table}

\begin{figure}
	\centering
		\includegraphics[scale=0.3]{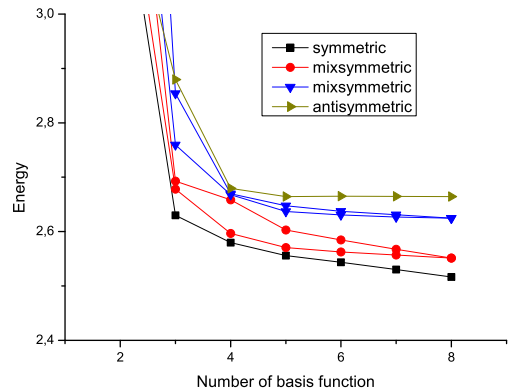}
	\caption{Convergence with basis set increase}%
	\label{fig:Graph2}
\end{figure}

\begin{figure}
	\centering
		\includegraphics[scale=0.3]{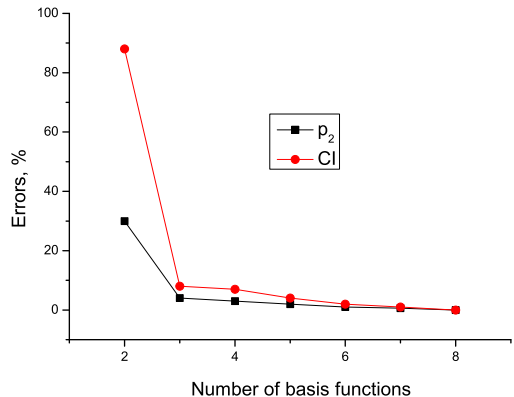}
	\caption{Errors with basis set increase}%
	\label{fig:Graph3}
\end{figure}
 
 The results show that for the complete basis of one-body functions $n_f=8$ the results 
of the applied methods of solution give exactly the same results. In all cases, the 
diagonalized matrices are of the same $n_f^3$ order providing the exact values for other
exited states. This situation continues in $p_t$ and $p_1$ up to $n_f=4$
for all states, and up to $n_f=3$ for the ground state in spite of one-body
basis set reduction. The result is a sequent that up to $n_f=4$ the number
of linear independent function constructed for $p_t$ and $p_1$ surfaces remains unchanged
and equals to 512. The ground state is symmetric and does not contains the exchange
hole, so correlations effect here is more important than for antisymmetric states.  Other 
states presented in Table II are mixed symmetric .                 

The decrease in the number of one-bode basis functions leads to the decrease in the order
of CI and $p_2$ matrices. As a result, the accuracy obtained with these methods drops 
significantly with the basis set reduction. This drop is shown in Fig. \ref{fig:Graph2} 
for the 6 lowest states of $p_2$ 
belonging to pure symmetric, mixed symmetric and pure antisymmetric states.  
As one can see in Fig. \ref{fig:Graph2} the loss of accuracy removes degeneration of the
mixed symmetric states. Fig. \ref{fig:Graph3} shows the relative errors of CI and 
$p_2$. Approximation $p_2$ gives about two times less errors in comparison with  
CI one.   

\section{He-like ions}
The He-like ions has been the subject of intensive study over last decades to
analyse the behavior of electrons in the nuclear field in its simplest two electron case
and learn how to construct the wave function for more complicated cases.
 
The are several types of wave functions used in the precise electronic structure calculations of He-like ions:
Hylleraas-type wave functions, conventional configuration interaction wave functions constructed from 
Slater-type orbitals, and configuration interaction wave functions with explicit dependence of the 
wave functions on $r_{12}$. In most calculations the Hylleraas-type wave function
\begin{equation}\label{Hyll}
	\Psi(\mathbf{r}_1,\mathbf{r}_2)=e^{-\alpha (r_1+r_2)}\sum_{i=1}^{l}c_i f_i(r_1,r_2,r_{12})
\end{equation}
where in pioneering works of Hylleraas\cite{hyller1,Hylleraas29} function $f_i(r_1,r_2,r_{12})$ has the form
\begin{equation}\label{H hy}
	f_i^H(r_1,r_2,r_{12})=(r_1+	r_2)^{l_i}(r_1-r_2)^{2m_i}r_{12}^{n_i}
\end{equation}
Frankowski and Peketis\cite{PhysRev.146.46} proposed another form for $f_i(r_1,r_2,r_{12})$
\begin{equation}\label{FP}
 \begin{split}
	f_i^{FP}(r_1,r_2,r_{12})=f_i^{H}(r_1,r_2,r_{12})\\
	\times [(r_1+r2)^2+(r_1-r2)^2]^{l_i/2}[ln(r_1+r_2)]^{k_i}
 \end{split} 
\end{equation}
multiplying $f_i^{H}$ by the logarithmic function. The double basis function method
with generalized Hylleraas functions
\begin{equation}
	\begin{split}
	\Psi(\mathbf{r}_1,\mathbf{r}_2)=e^{-\alpha_ar_1-\beta_ar_2}\sum_{ijk}c^a_{ijk}r_1^ir_2^jr_{12}^k\\
	+e^{-\alpha_br_1-\beta_br_2}\sum_{ijk}c^b_{ijk}r_1^ir_2^jr_{12}^k
	\end{split}
\end{equation}
was used in.\cite{DrakeZong} In works\cite{PhysRevA.15.1,PhysRevA.15.16,PhysRevA.38.26,Korobov}
the functions represented in the form
\begin{equation}
	\Psi(\mathbf{r}_1,\mathbf{r}_2)=(1+P_{12})\sum_ia_ie^{-\alpha_ir_1-\beta_ir_2-\gamma_ir_{12}}
\end{equation}
where $P_{12}$ is the operator permuting $r_1$ and $r_2$.
 These functions contain more than one nonlinear variational parameters 
in the exponent and up to several hundred coefficients $c_i$. 

All applications of Hylleraas-type wave functions to $He$ give the energy of the ground state
-2.9037236 a.e. and employment of more exact functions lead to the increase in the number of
significant decimal points.  

Conventional configuration interaction 
wave functions
\begin{equation}\label{cihe}
	\Psi(\mathbf{r}_1,\mathbf{r}_2)=\sum_ic_i\Phi_i(\mathbf{r}_1,\mathbf{r}_2),
\end{equation}
where $\Phi_i$ is a determinant function constructed from $i$ set of slater
spin-orbitals,  have been used for He-like ions calculations in Ref.\cite{Weiss}. 
An increase in the number of non-linear parameters allowed to reduce the expand 
length of configuration interaction wave function. 
Including an explicit dependence of a configuration interaction function on electron separation 
$r_{12}$ leads to a further decrease in the wave function expand length\cite{RootW,PhysRevA.15.1,Saha}.
Obviously, the application of the theory to solve the Schr\"{o}dinger equations for He-like ions has a 
particular importance . 

\subsection*{Equations.}
When solving the Schr\"{o}dinger equations for He-like ions, for length and energy it is convenient 
to use  the corresponding atomic units divided by nuclear charge $Z$ and $Z^2$,
respectively. In these units the Schr\"{o}dinger equation for $^1S_0$ state of He-like ions
can be written in the form
\begin{equation}
\begin{split}
H\Psi=\left[-\frac{1}{2}\left(\frac{\partial^2 }{\partial r_1^2}+
\frac{\partial^2 }{\partial r_2^2}\right)-\frac{1}{r_1}\frac{\partial }{\partial r_1}
-\frac{1}{r_2}\frac{\partial }{\partial r_2}\right.\\ \left.
-\frac{1}{r_1}-\frac{1}{r_2}+\frac{1}{Zp}\right]\Psi(r_1,r_2)=E\Psi(r_1,r_2)
\end{split}
\end{equation}
where $p=|\mathbf{r}_1-\mathbf{r}_2|$.
Let us represent a many-electron wave function in the form 
\begin{equation}\label{auxf}
\Psi(\mathbf{r}_1,\mathbf{r}_2)=\sum_i\chi_i(p)\Phi_i(\mathbf{r}_1,\mathbf{r}_2).
\end{equation}
where function $\Phi$ is a symmetrized production of two one-electron functions, 
and the weight function $\chi_i$ depends only on the total electron-electron interaction potential.
  
The set of equations for function $\chi_i$ can be written in the form
\begin{equation}\label{eqche}
\begin{split}
	\sum_j&\left[-\frac{t_{ij}(p)}{2}\frac{d^2\chi_j(p)}{dp^2}
	-\frac{1}{2}\left(\frac{2t_{ij}(p)}{p}+u_{ij}(p)\right)\frac{d\chi_j(p)}{dp}\right.\\ &\left.+
  \left(h_{ij}(p)+\frac{s_{ij}(p)}
{Zp}\right)\chi_j(p)\right]=E\sum_js_{ij}(p)\chi_j(p),
\end{split}
\end{equation}
with
\begin{align}
\label{S_he} s_{ij}(p)&=\langle\Phi_{i}(\mathbf{r}_1,\mathbf{r}_2)|
\Phi_j(\mathbf{r}_1,\mathbf{r}_2)\rangle_p\\
	 \label{T_he} t_{ij}(p)&=2s_{ij}(p), \\
	 \label{U_he} u_{ij}(p)&=\langle\Phi_i(\mathbf{r}_1,\mathbf{r}_2)
	|\sum_{k=1}^2\nabla_k p\nabla_k|
	   \Phi_j(\mathbf{r}_1,\mathbf{r}_2)\rangle_p\\
		&=\langle\Phi_i(\mathbf{r}_1,\mathbf{r}_2)\left|
		\frac{p^2+r_1^2-r_2^2}{2pr_1}\frac{\partial }{\partial r_1}\right.\\ 
		&+\left.\frac{p^2-r_1^2+r_2^2}{2pr_2}\frac{\partial }{\partial r_2}\right|
		\Phi_j(\mathbf{r}_1,\mathbf{r}_2)\rangle_p, \\
	 \label{H_he} h_{ij}(p)&=\langle\Phi_i(\mathbf{r}_1,\mathbf{r}_2
	|H_0(\mathbf{r}_1,\mathbf{r}_2)|
	\Phi_j(\mathbf{r}_1,\mathbf{r}_2)\rangle_p  ,
	\end{align}
	$H_0$ is the Hamilonian of non-interacting electrons and averaging over space coordinates is performed for constant particle separation $p$. 
	Representing $r_2=\sqrt{r_1^2+p^2-2r_1p\cos{\vartheta}}$ integration over sphere of
radius $p$ with the center at $r_1$ can be transformed into integration over $r_2$
\begin{equation}
	\int p\sin{\vartheta}d\vartheta d\phi=\frac{2\pi}{r_1p}\int_{|r_1-p|}^{r_1+p}r_2dr_2
\end{equation}
 and matrix elements of operator $\widehat{F}$ over the whole constant interaction 
potential was performed as
\begin{equation}
 \begin{split}
  F_{ij}(p)=\int dr_1 r_1^2\int_{|r_1-p|}^{r_1+p}\frac{r_2}{pr_1}dr_2
	\Phi_i(r_1,r_2)\\|F(r_1,r_2)|\Phi_j(r_1,r_2)	
 \end{split}
\end{equation}

	For description of $^1S_0$ states 
we will use $1s$, $2s$, $3s$ and $4s$ wave functions of an electron in the nuclear field 
written below 
\begin{align}
\phi_1(\mathbf{r})=2e^{-r}\nonumber\\
\phi_2(\mathbf{r})=	\frac{1}{\sqrt{2}}\left(1-\frac{r}{2}\right)e^{-r/2}\nonumber\\
\phi_3(\mathbf{r})=	\frac{2}{\sqrt{27}}\left(1-\frac{2r}{3}+\frac{2r^2}{27}\right)e^{-r/3}\\
\phi_4(\mathbf{r})=\frac{1}{4}\left(1-\frac{3r}{4}+\frac{r^2}{4}-\frac{r^3}{192}\right)e^{-r/4}	
\nonumber
\end{align}
From these functions four configurations with the lowest energies will be used
\begin{equation}\label{bfnc}
	\Phi_i(r_1,r_2)=[\phi_1(r_1)\phi_i(r_2)+\phi_1(r_2)\phi_i(r_1)]/\sqrt{2(\delta_{1i}+1)}
\end{equation}
Matrix elements between these functions on a surface $p$ can be obtained analytically 
and presented in Appendix where one can see that the expressions for matrix
elements between $\Phi_3$ and $\Phi_4$ functions contain very large numbers, 
and to avoid undesirable rounding errors the calculations were performed with 32 
significant numerals.

\begin{figure}
\begin{center}
\scalebox{0.30}{\includegraphics{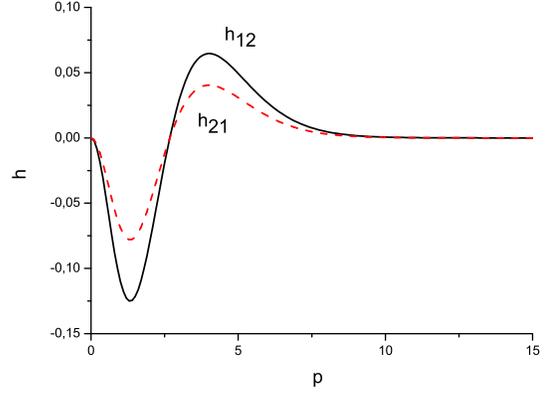}}
\caption{$H_{12}$ and $H_{21}$ matrix
 elements \label{f:h12} }
\end{center}
\end{figure}

\begin{figure}
\begin{center}
\scalebox{0.30}{\includegraphics{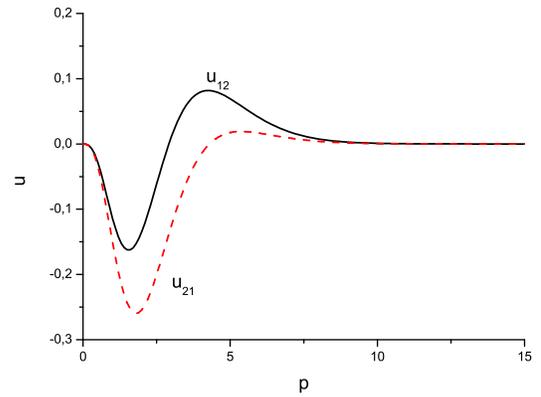}}
\caption{$U_{12}$ and $U_{21}$ matrix
 elements \label{f:u12} }
\end{center}
\end{figure}

\begin{figure}
\begin{center}
\scalebox{0.35}{\includegraphics{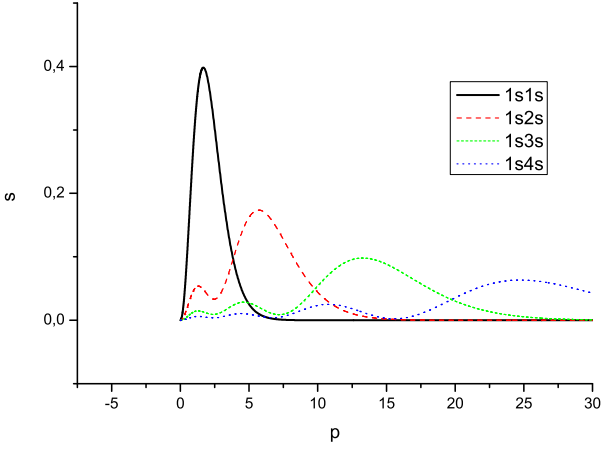}}
\caption{Diagonal overlap matrix elements.\label{f:1} }
\end{center}
\end{figure}

\begin{figure}
\begin{center}
\scalebox{0.35}{\includegraphics{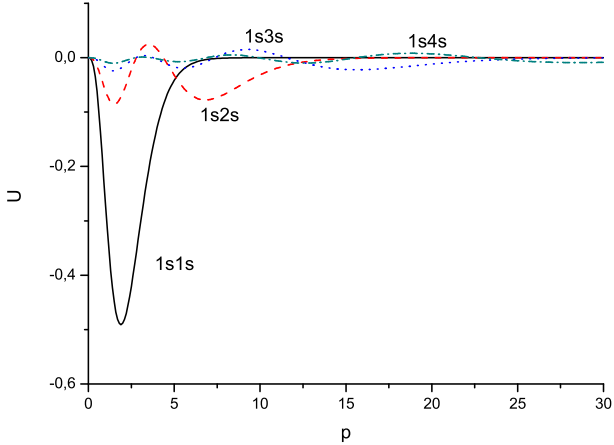}}
\caption{Diagonal U matrix elements.\label{u4}
}
\end{center}
\end{figure}

Overlap matrix $S(p)$ is symmetric, while matrices $H$ and $U$
are non-symmetric. Differences in matrix elements $H_{12}$, $H_{21}$ 
and $U_{12}$, $U_{21}$ are shown in 
Fig. \ref{f:h12} and Fig. \ref{f:u12} respectively.
The diagonal overlap matrix elements are shown in Fig. \ref{f:1}.
Note that orthogonal functions in the whole space (\ref{bfnc})
become non-orthogonal on the constant interaction potential surface. 
Off-diagonal matrix elements decay with $p$ growing faster then
the diagonal ones what has an important consequence - in $p\to \infty$
limit the set (\ref{eqc}) is split into independent equations.
The diagonal matrix elements of $U$ is
shown in Fig. \ref{u4}. One can see in Fig.\ref{f:1} and Fig.\ref{u4}
that both matrix elements $S$ and $U$ decay with $p$ growth, however not too
fast to neglect them in a precise calculations for rather big $p$.

The Hamiltonian matrix elements of non-interacting electrons reduced to
\begin{equation}\label{hij}
	h_{ij}=\epsilon_js_{ij}
\end{equation}
where $\epsilon_j$ is the energy of $\Phi_j$ state.
Due to $\epsilon_j$ factor $h_{ij}(p)\neq h_{ji}(p)$. Naturally, in the whole
space non-diagonal elements obtained by integration over $p$ equal to zero
and matrix $h$ become Hermitian.
 
\subsection*{Boundary conditions and computation details.}
 The boundary conditions 
for $\chi$ follow from the demand for $\Psi$ to be finite in the whole space
\begin{equation}\label{rstr}
\chi_i(p)s_{ij}(p)\chi_j(p)<\infty ,\ i,j=1,\ldots,n_f,
\end{equation}
$n_f$ is the size of basis set used. Series expansion of $U$ matrix elements shows
that $u_{ij}\to cp^3$ when $p\to 0$, i.e. $u_{ij}$ 
decay faster then $s_{ij}$ elements with $p\to 0$.
Thus, for small $p$ Eqs.(\ref{eqche}) can be approximated by equation
\begin{equation}
	2\frac{d\omega}{dp}+\frac{\omega}{Z}=0
\end{equation}
with solution 
\begin{equation}
	\omega(p)=e^{Zp/2}
\end{equation}
Determine the general solutions  of (\ref{eqche}) as
\begin{equation}
	\chi^l(p,E)=\sum_i c^l_i\chi_i^l(p,E)
\end{equation}
where $c^l_i$ are arbitrary constants and $\chi_i^l(p,E)$ is partial
solution of (\ref{eqche}) with the initial values
\begin{align}\label{brd0}
	\chi^l_i(0)&=\omega(0)\sum_j\delta_{ij}\\ \label{dbrd0}
	\frac{d\chi^l_i(0)}{dp}&=\frac{Z}{2}\chi^l_i(0)
\end{align}
The superscript $l$ means that $\chi^l$ are determined from the left
initial values, each $\chi^l_i$ is a vector function.   

To defined the boundary condition when $p\to\infty$, we represent
the solution of (\ref{eqche}) at a point $p$ as $e^{\lambda p}$. 
Substitution of this representation in (\ref{eqche}) leads to
\begin{equation}\label{eqphe}
\begin{split}
	\sum_{j=1}^n\left[-\lambda^2s_{ij}(p)
	-\lambda ((2s_{ij}(p)/p+u_{ij}(p))+h_{ij}(p)\right.\\ \left.
	+s_{ij}(p)/Zp-Es_{ij}(p)\right]\chi_j(p)=0,\ i=1,...,n_f,
\end{split}
\end{equation}
 Set (\ref{eqphe}) has non-zero solution if
\begin{equation}\label{detihe}
\det(\Lambda)	=0
\end{equation}
where matrix $\Lambda$ is determined by the expressions in the square brackets
of (\ref{eqphe}). 
Obviously, $\det(\Lambda)$ is a $2n_f$ order polynomial of $\lambda$ the
$2n_f$ roots of which, possibly complex, will be denoted $\lambda_i$. Not all of
the roots  satisfy the condition (\ref{rstr}). In particular, for $n_f=1$ taking 
into account that $u_{11}/s_{11}\to -2$ when $p\to \infty$, the two roots of 
(\ref{detihe}) in this limits are
\begin{equation}
	\lambda_{1,2}=1\pm\sqrt{1+\epsilon_1-E}
\end{equation}
Function $\chi$ satisfies to the condition (\ref{rstr}) only for $\lambda_2$. It is a growing function
for $E>\epsilon_1$ and a decreasing one for $E<\epsilon_1$. For $E=\epsilon_1$ $\lambda_2=0$ and
the configuration weight function becomes a constant as it should be for non-interaction electrons.

\begin{figure}
\begin{center}
\scalebox{0.35}{\includegraphics{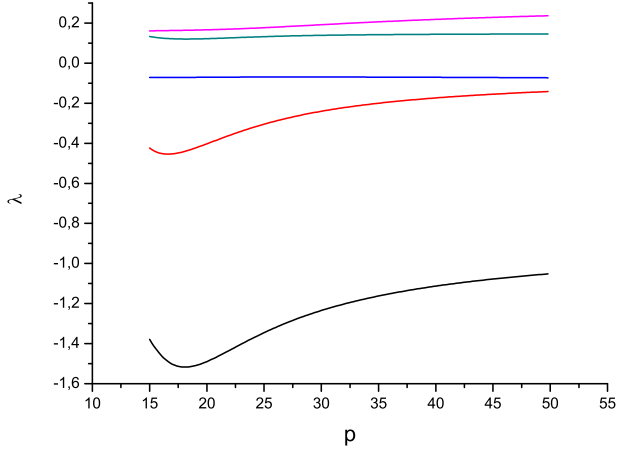}}
\caption{Roots of (\ref{deti}) satisfying to (\ref{rstr}) as a function
of $p$ for $n_f=4$. Physical meaning
have the four lowest roots. \label{f:root}
}
\end{center}
\end{figure}
For $n_f=4$ the dependence on $p$ of the roots satisfying condition (\ref{rstr}) 
is shown in Fig.\ref{f:root}. Determine partial solutions of (\ref{eqc})
 $\chi_i^r(p,E)$ which satisfy boundary conditions
\begin{equation}\label{brdp}
\begin{split}
	\chi^r_i(p,E)&=e^{\lambda p}\\
	\frac{d\chi_i^r(p,E)}{dp}&=\lambda_i\chi^r_i(p,E)
\end{split}
\end{equation}
The general solution of (\ref{eqc}) with these right boundary conditions
can be presented in the form
\begin{equation}
	\chi^r(p,E)=\sum_{i}^n c_i^r\chi_i^r(p,E)
\end{equation}
Coefficients $c^l$ and $c^r$ are determined from the demand that functions 
$\chi_i^l$ must continuously pass to functions $\chi_i^r$ at a point $p$
together with their 1st derivatives 
\begin{align}
	\chi_i^l(p,E)=\chi_i^r(p,E)\\
	\frac{\chi_i^l(p,E)}{dp}=\frac{\chi_i^r(p,E)}{dp}
\end{align}
 For solubility of this set of equation it is necessary that
 \begin{equation}\label{detm}
\det(\Omega(E))=0	 
 \end{equation}
where
\begin{equation}
	\Omega=\left|\begin{array}{cc}
	  \chi^l(p,E) & -\chi^r(p,E)\\
		\frac{\chi^l(p,E)}{dp} & -\frac{\chi^r(p,E)}{dp}
		\end{array} 
		\right|
\end{equation}
Condition (\ref{detm}) determines the energy $E$ of the system.

To solve (\ref{eqche}) for $n_f<4$ with the bounder condition (\ref{brd0}) or (\ref{brdp})
the Runge-Kutta 4th-order method can be employed. For $n_f=4$ equations (\ref{eqche})
become stiff and it is impossible to obtain solution with Runge-Kutta method
due to fast growth of rounding errors leads to divergence of the searching solution. 
Described in\cite{Rosenb} Rosenbrock method elaborated for stiff equations also failed 
to solve the problem. We succeed in solving (\ref{eqche}) exploiting tridiagonal matrix 
algorithm (Thomas algorithm)\cite{Yanenko}. For these equations (\ref{eqche}) were 
approximated with
\begin{equation}\label{deqc}
	\mathbf{A}_k\chi(k-1)+\mathbf{B}_k\chi(k)+\mathbf{C}_k\chi(k+1)=0
\end{equation}
Here $k$ numerates points of $p$-mash, 
\begin{align} 
	\mathbf{A}_k&=-2\mathbf{t}_k/d^2+(\mathbf{h}_k/p_k+
	\mathbf{u}_k/2)/d\\
	\mathbf{C}_k&=-2\mathbf{t}_k/d^2-(\mathbf{s}_k/p_k+
	\mathbf{u}_k/2)/d\\
	\mathbf{B}_k&=\mathbf{h}_k+(q/p_k-E)\mathbf{s}_k
\end{align}
$d=p_{k+1}-p_k$, $\mathbf{t}$, $\mathbf{u}$, $\mathbf{h}$ and $\mathbf{s}$ are 
$n_f\times n_f$ matrices, $\chi$ is $n_f$-order vector.

In line with the Tomas algorithm partial solutions of (\ref{deqc}) can be represented as
\begin{equation}\label{bwl}
	\chi_i(k)=\mathbf{X}^l_{ik}\chi(k+1),\; i=1,\ldots,n_f
\end{equation}
where matrix
\begin{equation}\label{upl}
	\mathbf{X}^l_{ik}=-(\mathbf{A}_k\mathbf{X}^l_{i,k-1}+\mathbf{B}_k)^{-1}\mathbf{C}_{k}
\end{equation}
with
\begin{equation}\label{x0}
 \mathbf{X}^l_{i,0}=e^{-Zd/2}\delta_{ij},\;j=1,\ldots,n_f	
\end{equation}
 The choice of $\mathbf{X}^l_0$ follows from (\ref{dbrd0}) and determines the
correct 1st derivatives of the function rather than the function values. Thus the 
solution of (\ref{deqc}) is through the calculation of $\mathbf{X}^l$ with (\ref{upl}) 
in upward direction at the first stage and the calculation of $\chi$ with (\ref{bwl}) 
in backward direction. 

Obviously the algorithm can be reversed, i.e. the calculation of $\mathbf{X}^r$ starting 
from a big $p$ and calculate $\chi$ in the backward direction. 
Corresponding formulas for a partial solution are presented below
\begin{align}\label{upr}
	\chi^r_i(k+1)&=\mathbf{X}^r_{ik}\chi_i(k),\;i=1,\ldots,n_f\\ \label{bwr}
\mathbf{X}^r_{i;k-1}&=(\mathbf{A}_k\mathbf{X}^r_{ik}+\mathbf{B}_k)^{-1}\mathbf{C}_k\\ \label{xr}
 \mathbf{X}^r_{in}&=e^{\lambda_i d}\delta_{ij}\;i=1,\ldots,n_f	
\end{align}
$\lambda_i$ are roots of (\ref{detihe}). (\ref{xr}) provides bounder conditions for a partial solution 
of (\ref{deqc}) for big $p$. 

In principle, one can use to solve (\ref{deqc}) formulas (\ref{bwl})-(\ref{x0}) or 
(\ref{upr})-(\ref{xr}). However, computational errors can grow with moving off the border. 
To decrease these errors, it is useful to apply both of these ways, matching their solution 
at some point inside $p$-interval. At this point functions $\chi_i$ and their 1st 
derivatives calculated with $\mathbf{X}^l$ and $\mathbf{X}^r$
must be equal to each other. The derivatives can be presented in the forms
\begin{align}\label{derivl}
	\chi'_i(m)=(\mathbf{I}-\mathbf{X}^l_{im})\chi_i(m)/d\\ \label{derivr}
	\chi'_i(m)=(-\mathbf{I}+\mathbf{X}^r_{im})\chi_i(m)/d
\end{align}
The matching conditions lead to a set of equations 
\begin{equation}\label{beqc}
\begin{split}
	\sum_i\chi^l_i(m)c^l_i&=\sum_i\chi^r_i(m)c^r_i\\
	\sum_i(\mathbf{I}-\mathbf{X}^l_{i,m-1})c_i^l&=
	\sum_i(\mathbf{I}-\mathbf{X}^r_{i,m+1})\chi_i(m)c^r_i
\end{split}
\end{equation}
The set of equations (\ref{beqc}) determined the system energy because the
set has nonzero solution only for selected energies making the
determinant of the set equals to zero. As seen in Fig.\ref{f:root} $\lambda_i(p)$
tends to constant when $p\rightarrow \infty$ and a use of finite $p$ introduce an
errors in to the calculated energy. From the other side, the numerical errors tends
to grows for too large $p$.  In energy calculations we used $p=40$. This value is a 
compromise between the variation of $\lambda_4$ and the increasing numerical errors 
with $p$ growth.

\begin{table*}[tb]
\vspace{5mm}
TABLE I. The ground states energies of He-like ions. 
\begin{center}
\begin{tabular}{ccccccccc} \hline \hline
Ion&\multicolumn{7}{c}{ Energy, a.u.} \\
     &HF$^a$ &1&2&3&4& CI$^b$ & Hyl$^c$& Exp.$^d$  \\ 
   \hline
$H^{-}$    &          &  -0.498461 &  -0.526779 &  -0.527133&  -0.527790& -0.5277303&& \\
He         &- 2.86171 &  -2.879388 &  -2.900539 &  -2.902257&  -2.903756& -2.9037236& -2.903724& -2.90338\\                    
$Li^{+}$   & -7.23633 &  -7.256393 &  -7.276105 &  -7.278158&  -7.279468& -7.279819 & -7.279913& -7.278956\\                  
$Be^{2+}$  &-13.61130 & -13.632404 & -13.651487 & -13.653685& -13.655578& -13.655551&-13.655566& -13.6574\\                 
$B^{3+}$   &-21.98607 & -22.008016 & -22.026751 & -22.029031& -22.031332& -22.030875&-22.030972& -22.0360\\                   
$C^{4+}$   &-32.36137 & -32.383429 & -32.401946 & -32.404281& -32.407322& -32.406070&-32.406247&  -32.4174\\   
$N^{5+}$   &-44.73618 & -44.758728 & -44.777098 & -44.779475& -44.781458& -44.781141&-44.781445& -44.8035\\ 
$O^{6+}$   &-59.11159 & -59.133956 & -59.152223 & -59.154631& -59.156576& -59.156222&-59.156595& -59.1958\\ 
$F^{7+}$   &-75.48702 & -75.509136 & -75.527329 & -75.529764& -75.532249& -75.531401&-75.531712& -75.54413   \\
$Ne^{8+}$  &-93.86174 & -93.884283 & -93.902421 & -93.904878& -93.910240& -93.906452&-93.906807& -94.0086\\  
$Na^{9+}$  &          &-114.259406 &-114.277503 &-114.279981&-114.283217&-114.28165 &&\\
$Mg^{10+}$ &          &-136.634511 &-136.652577 &-136.655073&-136.659456&-136.65672 &&\\                    
$Al^{11+}$ &          &-161.009602 &-161.027646 &-161.030158&-161.044494&-161.03180 &&\\                  
$Si^{12+}$ &          &-187.384681 &-187.402709 &-187.405237&-187.412848&-187.40687 &&\\                 
$P^{13+}$  &          &-215.759753 &-215.777769 &-215.780312&-215.78715 &-215.78191 &&\\                   
$S^{14+}$  &          &-246.134816 &-246.152826 &-246.155383&-246.159333&-246.15697 &&\\   
$Cl^{15+}$ &          &-278.509875 &-278.527880 &-278.530450&-278.535628&-278.53201 &&\\ 
$Ar^{16+}$ &          &-312.884928 &-312.902932 &-312.905515&-312.913206&-312.90704 &&\\ 
\hline \hline                                     
$^a$ Ref.\cite{Clementi}. \\  
$^b$ Ref.\cite{Saha}. \\  
$^c$ Ref.\cite{PhysRevA.50.854}  \\                                 
$^d$ Ref.\cite{Moore,Bashkin}                                   
\end{tabular}                                      
\end{center}  
\end{table*}   

\subsection*{Results.}
The energies obtained with (\ref{eqche}) for the ground states of He-like ions are 
presented in Table I together with HF and configuration interaction results.
The use of only one configuration in (\ref{auxf}) gives energies slightly below 
Hartree-Fock limit. Inclusion 2nd and 3rd configurations gives the results comparable
but slightly above those of CI with 35 configurations. When the fourth configuration 
is added, the energies fall below the CI results and below Hylleraas limit  
excepting of $O^{6}$. 

\begin{figure}
\label{wf1}
\begin{center}
\scalebox{0.35}{\includegraphics{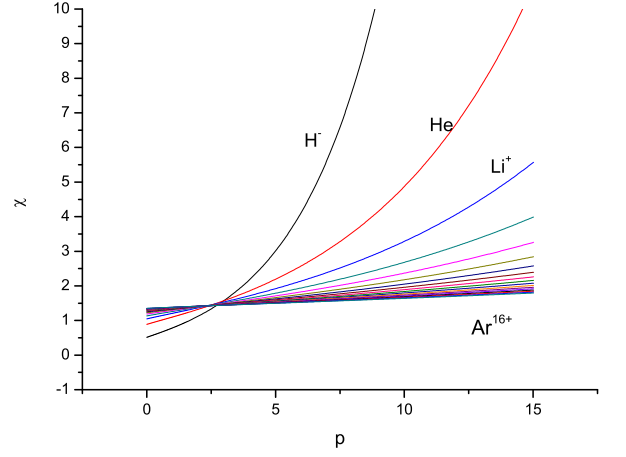}}
\caption{Configuration wave functions of $H^-$,...,$Ar^{16+}$ for $1s1s$ 
        configuration approximation.\label{f:wf1}}
\end{center}
\end{figure}

The configuration weight functions $H^-,\ldots ,Ar^{16+}$ for one-configuration approximation
are shown in Fig.\ref{f:wf1}.  The $H^-$ configuration weight function demonstrates the most rapid
growth with $p$. The functions growth slow down with the increase in nuclear charges and 
tends to a constant, demonstrating a relative decrease in electron-electron interaction as
compared to the nuclear field. The growing interaction function decreases the probability 
to find electron at a small separation and increase at a bigger separation in comparison with
non-interacting cases.

\begin{figure}
\begin{center}
\scalebox{0.35}{\includegraphics{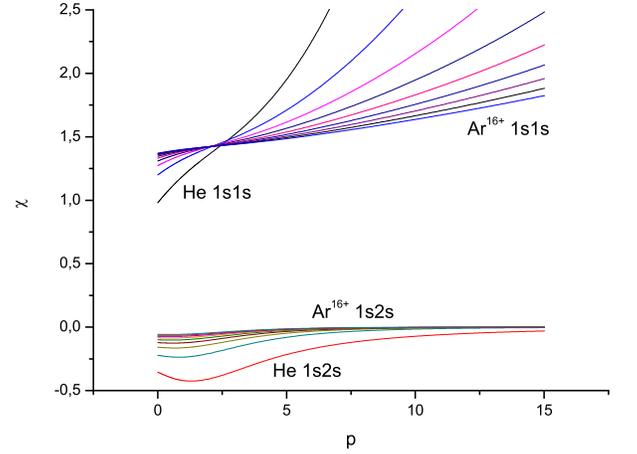}}
\caption{$1s1s$ and $1s2s$ configurations wave functions
from $He$ to $Ar^{16+}$ for even atomic numbers 
for 2-configuration approximation.
\label{f:wf2}} 
\end{center}
\end{figure}

The configuration weight functions from $He$ to $Ar^{16+}$ for even atomic numbers are
shown in Fig.\ref{f:wf2}. As one can see $1s1s$ configuration weights are similar to the 
configuration weight functions for $n_f=1$ (see Fig.\ref{f:wf1}). The $1s2s$ functions have 
noticeable values for small $p$ which tends decrease with the growth of $p$ and 
the atomic number. 

\begin{figure}
\begin{center}
\scalebox{0.35}{\includegraphics{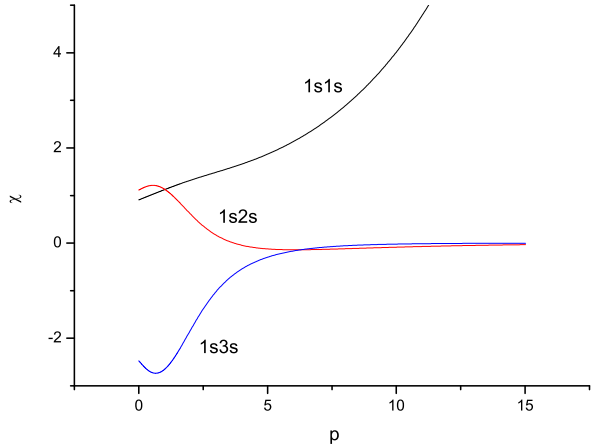}}
\caption{$He$ configuration wave function 
 for $n_f=3$. \label{f:wf3He}}
\end{center}
\end{figure}

\begin{figure}
\begin{center}
\scalebox{0.35}{\includegraphics{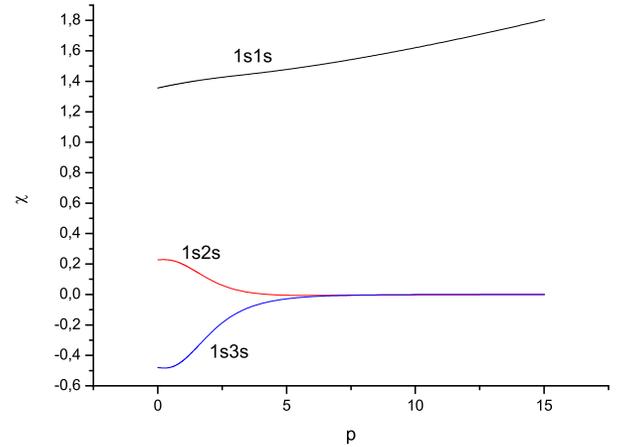}}
\caption{$Ar^{16+}$ configuration wave functions 
          for $n_f=3$. \label{f:wf3Ar}}
\end{center}
\end{figure}

The configuration weight functions of $He$ and $Ar^{16+}$  
for 3-configuration approximation are shown in Fig.\ref{f:wf3He} and Fig.\ref{f:wf3Ar}, correspondingly. 
\begin{figure}[t]
\begin{center}
\scalebox{0.35}{\includegraphics{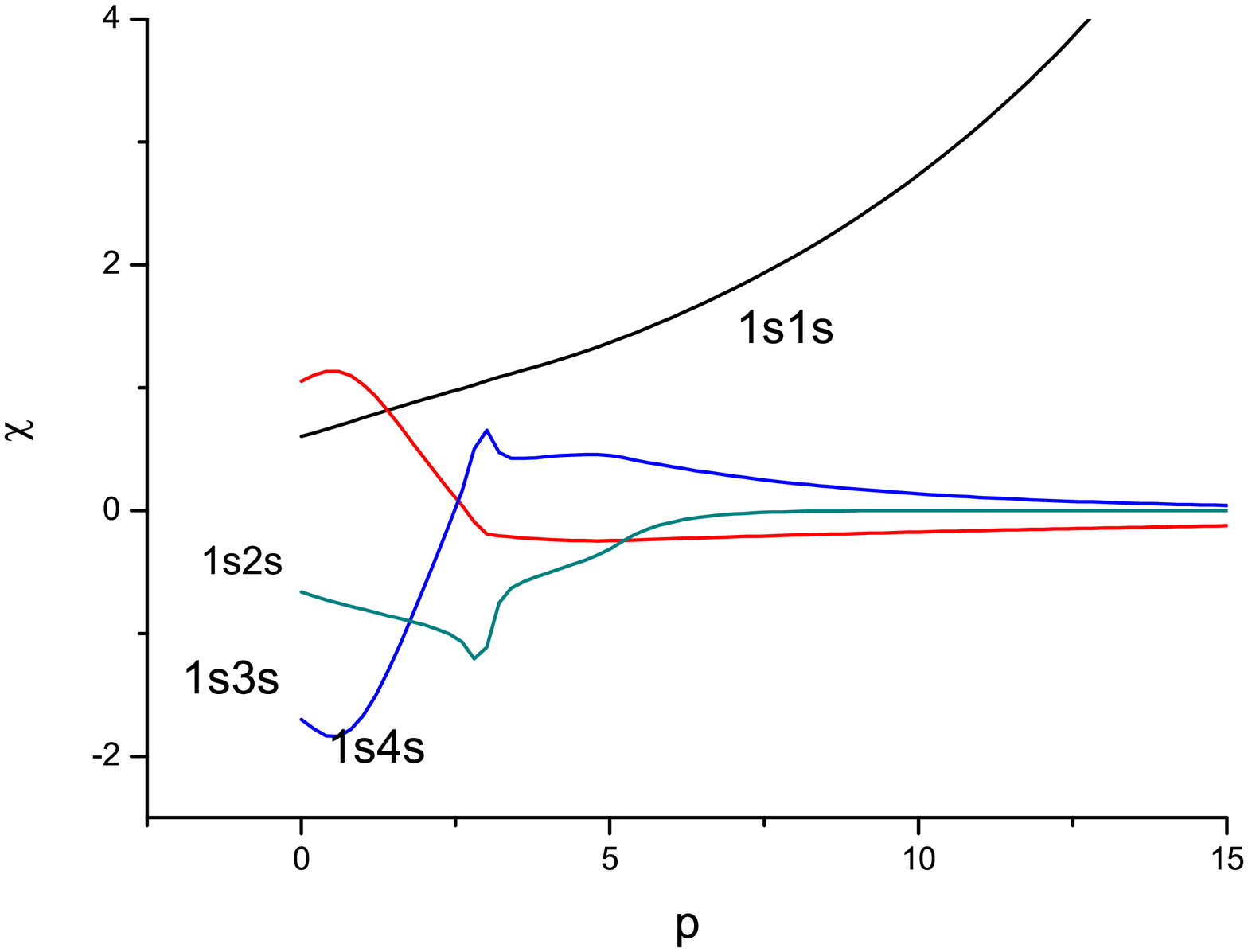}}
\caption{$He$ configuration wave functions for $n_f=4$. \label{f:wf4}}
\end{center}
\end{figure}
It can be seen that the absolute value of $1s3s$ weight function for $He$ in small
$p$ region significantly exceeds the approximately equal contributions of $1s1s$
and $1s2s$ configurations, with the growth of $p$ contribution $1s1s$ dominating.
When nuclear charge increases the contributions of $1s2s$ and $1s3s$ configurations
tend to decrease. The weight functions for $Ar^{16+}$ are similar to those for $He$;
however, the relative contribution of $1s2s$ and $1s3s$ to the wave function decreases
in comparison with $1s1s$ contribution.

The configuration weight functions for $He$ with $n_f=4$ are shown Fig.\ref{f:wf4}.
The contribution of $1s1s$, $1s2s$ and $1s3s$ interact weight function into the 
wave function are similar $n_f=3$ case; however, a peak and a visible knee close to 
$p=2.6$ appear at the $1s2s$ and $1s3s$ weight functions. The absolute value of $1s4s$  
contribution is comparable with $1s1s$ contributions and reaches a maximum close to 
$p=2.6$ and then drops down.

\section{Conclusions}
The proposed theory can be considered as an extension of configuration 
interaction method in which contributions of different configurations to
the wave function become dependent on the values of interaction potential, 
which makes the wave function more flexible and eliminates the influence of 
the wave function cusps on the convergence of the wave function to the exact 
one with a basis set increase. From the other side, the theory can be compared 
with explicitly correlated $R12$ and $F12$ methods since coefficients of wave 
function expansion over configurations depend on the inter particle separations 
and can be considered as a kind of wave function factors explicitly depending 
on a particle-particle separation. The main difference between these theories 
is the form of dependence of these factors on particle-particle separation which, 
in explicitly correlated theories, is prescribed whereas in the presented theory 
the factors are obtained by the solution of the corresponding weight function 
equations (\ref{eqc}).

Equations (\ref{eqc}) were developed by energy variation,  
therefor, they provide upper bounds to the ground-state energy.

The important future of the proposed method,  as opposed to common methods of 
electronic structure calculations, is employing a basis set of non-interacting 
particle which does not presupposed the use of iteration procedure of Hartree-Fock method. 

The solution of model examples proves that the theory is correct. 
The energies obtained with approximations to the theory are grater than the exact 
ones and converged to the exact results, so these approximations satisfy 
the variational principle. The convergence of CWF method with basis set increase even 
in its lowest approximation is faster than that of CI method. 

The performed calculations show that the developed theory in the lowest approximation
with only one configuration of non-interacting particles gives energies
of He-like ions below the Hartree-Fock limits.  The use of three configurations 
constructed from $1s$, $2s$, and $3s$ wave functions 
of non-interacting electrons in the nuclear field gives ground state energies of 
He-like ions close to configuration interaction wave function with 35 configurations 
constructed from seven $s$, $p$, $d$, $f$, and $g$ Slater type orbitals and with 
configuration interaction wave function with 15 configuration constructed from 5 Slater 
orbitals and explicit $r_{12}$ terms up to the 5th order. Addition of the 4th 
configuration with $4s$ functions gives the energies below the CI method and 
the Hylleraas limit.  The results were obtained without iteration procedure of 
self-consistent field because the developed theory does not presuppose the use of 
the Hartree-Fock approximation as a preliminary step for precise calculations.

The equations (\ref{eqc}) were obtained by energy variation and their application to the 
solution of the simple modelsshows that such equations do not contradict
the variational principle, so the reasons why the obtained energies with $n_f=4$ turn out 
to drop below the most precise calculations should be sought elsewhere. Most 
probably the numerical calculations has been performed with insufficient accuracy. 
We used direct numerical solutions of (\ref{eqc}). If for $n_f<4$ the application of the
Runge-Kutta algorithm makes it possible to perform the calculations with a given accuracy,
whereas for $n_f=4$ this algorithm does no work since the equations become too stiff and 
numerical errors become unacceptable. Moreover, the application of the Rosenbrock method 
for solving stiff equations also failed to solve the problem. The Thomas algorithm used 
in this work significantly reduced the numerical errors, however it needs improving to 
guarantee the desired accuracy. Another way to solve (\ref{eqc}) is to search the solutions 
in the form of a linear combination of some basis functions as it was done in all precise 
methods. In this case one has to find a basis which will be complete and fast converging.  
  
It should be noted that any expansion of the theory on many-atomic systems presupposes 
the construction of molecular orbitals of non-interacting electrons. It is these orbitals 
that should be used in averaging of one-bode operators over interaction potential surfaces, 
whereas surfaces themselves do not depend on nuclear positions.

\vspace{\baselineskip}
\section*{Acknowledgments}
 
 The author gratefully acknowledges
helpful discussions with the colleagues from Laboratory of Quantum Chemical of Boreskov
Institute of Catalysis. 
  
\bibliographystyle{aipnum4-1}
%
\end{document}